\documentclass[11pt]{article}
\usepackage{graphicx}

\title{$K^0$ meson physics in the gravitation field: a constraint on
the equivalence principle}

\author{Savely G. Karshenboim\\
~\\
Max-Planck-Institut f\"ur Quantenoptik, Garching, 85748, Germany\\
D.~I. Mendeleev Institute for Metrology, St.Petersburg, 190005,
Russia\\
{\em email\/}: savely.karshenboim@mpq.mpg.de}

\begin{document}

\maketitle

\begin{abstract}
$K^0$--$\overline{K^0}$ oscillations are extremely sensitive to the
$K^0$ and $\overline{K^0}$ energy at rest. Even assuming
$m_{K^0}=m_{\overline{K^0}}$, the energy is not granted to be the
same if gravitational effects on $K^0$ and $\overline{K^0}$ slightly
differ. We consider various gravitation fields present and, in
particular, galactic fields, which provide a negligible
acceleration, but relatively large gravitational potential energy. A
constraint from a possible effect of this potential energy on the
kaon oscillations is found to be
\[
\left\vert\left(\frac{m_g}{m_i}\right)_{K^0}-\left(\frac{m_g}{m_i}\right)_{\overline{K^0}}
\right\vert\leq 8\times 10^{-13}~~~{\mbox{at~CL=90\%}}\;.
\]
The derived constraint is competitive with other tests of
universality of the free fall. Other applications are also
discussed.
\end{abstract}

\section*{Introduction}

General relativity suggests the equivalence principle, which means
that the gravitational mass is equal to the inertial one, or, which
is the same, that the acceleration of free fall is universal. A
number of tests for the universality of the free fall acceleration
have reached an impressive level of below a part in $10^{12}$.
In particular, a laboratory test was performed on beryllium and
titanium test bodies in the Earth gravitational field
\cite{equivalence}
\begin{equation}
\eta_{\rm Be-Ti}=\Bigl(0.3\pm1.8\Bigr)\times 10^{-13}\;,
\end{equation}
and evaluation of the Lunar Laser Ranging data delivered a
constraint on the Earth-Moon system in the Sun field \cite{turyshev}
\begin{equation}
\widetilde{\eta}_{\rm E-M}=\Bigl(-1.0\pm1.4\Bigr)\times 10^{-13}\;,
\end{equation}
 where
\begin{eqnarray}
\eta_{ab}&=&\frac{2(g_a-g_b)}{g_a+g_b}\;,\nonumber\\
\widetilde{\eta}_{ab}&=&\left(\frac{m_g}{m_i}\right)_{a}-\left(\frac{m_g}{m_i}\right)_{b}\;,
\end{eqnarray}
$g_{a,b}$ is the acceleration of free fall of the body $a$ or $b$
towards the Earth or the Sun, and $m_i$ and $m_g$ are the inertial
and gravitational masses, correspondingly, of the probe body  $a$ or
$b$.

Parameters $\eta_{ab}$ and $\widetilde{\eta}$ are very similar to
each other and parameterize a deviation from the equivalence
principle. Under certain assumptions they coincide.

Most of the paper is devoted to the system $K^0$--$\overline{K^0}$
in the galactic gravitational field, while in the conclusions we
consider various other probe particles and existing fields.
Eventually we derive a constraint on the value of
$\widetilde{\eta}_{{K^0}\overline{K^0}}$.

Still, there is a certain difference in its interpretation comparing
with a traditional constraint for  $\eta$ and $\widetilde{\eta}$.
Usually, we know from independent experiments that $m_g\approx m_i$.
Such a statement for kaons may be obtained as a result of
model-dependent indirect interpretation of existing data.

\section*{Basics}

Our consideration includes the following elements
\begin{itemize}
\item we assume that in the absence of gravity there is CPT symmetry
and $m_{K^0}=m_{\overline{K^0}}$ (looking for any exotic physics,
such as possible violation of the universality of the free fall one
cannot blindly rely on an established paradigm, but rather state
explicitly, which theoretical suggestions and experimental data are
involved);
\item we suggest that the appropriate approximation for
gravitation-induced corrections to the $K^0$--$\overline{K^0}$
oscillations is consideration at zero velocity;
\item we suggest that the two `local' established components of
motion of the Solar system with velocities of 259 km/s and 185 km/s
\cite{galaxy} can be treated as a kind of circular motion on a
Newtonian orbit.
\end{itemize}

The idea of our constraint is very simple. Experimental
consideration of the ${K^0}$--$\overline{K^0}$ oscillations sets a
constraint on the value of (see, e.g., \cite{pdg,kaon})
\begin{equation}
\frac{\left\vert m_{K^0}-m_{\overline{K^0}}\right\vert}{m_{K^0}}\leq
0.8\times 10^{-18}~~~{\mbox{at~CL=90\%}}\;,
\end{equation}
which is valid, however, once we assume the equivalence principle.
In general, the quantity in the left-hand side of the inequality is
rather to be substituted for
\begin{equation}
\frac{\left\vert
E^{(0)}_K-E^{(0)}_{\overline{{K^0}}}\right\vert}{m_Kc^2}\;,
\end{equation}
where $E^{(0)}$ is the energy at rest, which includes the
gravitation energy. We believe that for slow-meson experiments a
consideration of oscillations at rest is a good approximation which
takes into account a main effect of gravity within or beyond the
equivalence principle.

As long as the equivalence principle is assumed, the gravitation
effects for kaon and antikaon cancel in the numerator and, as long
as CPT symmetry is assumed, the non-gravitational effects also
cancel. Otherwise we obtain:
\begin{equation}
\frac{\left\vert
E^{(0)}_K-E^{(0)}_{\overline{{K^0}}}\right\vert}{m_Kc^2}=
\left\vert
\left(\frac{m_g}{m_i}\right)_{K^0}-\left(\frac{m_g}{m_i}\right)_{\overline{K^0}}
\right\vert
\; \left\vert\frac{U(r)-U(\infty)}{c^2}\right\vert \;,
\end{equation}
where because of CPT invariance we assume that at $r=\infty$ the
numerator is equal to zero and that its deviation from zero can only
be caused by the non-universality in the gravitational interaction,
which in particular may be not the same for ${K^0}$ and
$\overline{K^0}$.

\section*{Galactic gravitation fields}

The motion of the Solar system in respect to the frame where the
cosmic microwave background radiation is isotropic has at least
three components and two of them with velocities  259 km/s and 185
km/s can be identified \cite{galaxy} as related to rotation around
certain galactic-scale attractors.

For a rotation around a center, which is free fall, we can write for
gravitational acceleration
\begin{eqnarray}
a &=& - \frac{U(r)}{r}\nonumber\\
&=&\frac{{\bf v}^2}{r}
\end{eqnarray}
and find the potential $U(r)$, defined so that $U(r=\infty)=0$. In
this case, we need to know only the velocity of rotation
\[
U(r)=- {\bf v}^2\;.
\]
If we have a few attractors, then
\[
U_{\rm tot}=\sum_i U_i\;.
\]
Summing the two well-understood rotating components mentioned above
\cite{galaxy} we arrive at
\[
-\frac{U_{\rm tot}}{c^2}=\frac{\sum_i { {\bf
v}_i^2}}{c^2}\simeq10^{-6}\;,
\]
as follows from their velocities.

That leads us to a constraint on the equivalence of the free fall
acceleration for kaon and antikaon
\begin{equation}
\bigl\vert\widetilde{\eta}_{{K^0}\overline{K^0}}\bigr\vert\leq
8\times 10^{-13}~~~{\mbox{at~CL=90\%}}\;,
\end{equation}
which is compatible with other tests \cite{equivalence,turyshev}.

We have ignored the largest velocity component of 455 km/s
\cite{galaxy} since it is unclear whether it may be attributed to a
certain rotation. However, any additional rotational component will
only increase $U_{\rm tot}$ and thus will make the constraint even
stronger.

This test is very specific for three reasons:
\begin{itemize}
\item It includes a theoretical assumption (such as CPT invariance: $m_{K^0}=m_{\overline{K^0}}$).
\item The probe particles are not baryons.
\item The field source is not purely baryonic and includes a portion of the dark matter.
\end{itemize}
Meanwhile, the other precision tests \cite{equivalence,turyshev} do
not need CPT invariance for evaluation of the data, they are
performed either on bulk objects or on atomic beams in a field from
a pure baryonic-matter source.

\section*{Other gravitational sources and other probe particles}

One can perform less accurate tests studying
${D^0}$--${\overline{D^0}}$ and ${B^0}$--${\overline{B^0}}$
oscillations. From \cite{pdg} we estimate
\begin{eqnarray}
\frac{\bigl\vert m_{D^0}-m_{\overline{D^0}}\bigr\vert}{m_{D^0}}\leq
10^{-15}\;,\nonumber\\
\frac{\bigl\vert m_{B^0}-m_{\overline{B^0}}\bigr\vert}{m_{B^0}}\leq
10^{-12}\;.\nonumber
\end{eqnarray}

That sets still strong constraints on some exotic components of
gravity at the level of
\begin{eqnarray}
\bigl\vert\widetilde{\eta}_{{D^0}\overline{D^0}}\bigr\vert\leq
10^{-9}\;,\nonumber\\
\bigl\vert\widetilde{\eta}_{{B^0}\overline{B^0}}\bigr\vert\leq
10^{-6}\;.\nonumber
\end{eqnarray}

If one considers the galactic gravitational attractors as not well
established, we may consider the rotation of Earth in respect to the
Sun with the velocity of 30 km/s which means
\[
\frac{U_{\rm Sun}(r_\infty)-U_{\rm Sun}(r_{\rm Earth})}{c^2}\simeq
10^{-8}\;.
\]
That makes all constraints two orders of magnitude weaker.

\section*{Antigravity for antiparticle}

As a kind of ultimate violation of the equivalence principle, which
may also involve CPT violation, there is a suggestion for
antigravity of antimatter. In this case we can consider for these
experiments a completely different approach to field sources and we
can also abandon an {\em a priori\/} assumption that the masses of
neutral mesons and their antiparticles are the same.

We note that the difference in the Sun-Earth distance for the
perihelion and aphelion of the Earth orbit is sufficient to maintain
an observable potential difference
\begin{equation}
\frac{U_{\rm Sun}(r_{\rm perihelion})-U_{\rm Sun}(r_{\rm
aphelion})}{c^2}\simeq 3.2\times10^{-10}\;,
\end{equation}
which allows to rule out the antigravity for all three kinds of
neutral mesons considered above. In principle, mesons consist of a
quark and an antiquark ($K^0=\overline{s}d$, $D^0= c\overline{u}$,
$B^0=\overline{b}d$), however, the difference in masses of the quark
and antiquark is quite large and any model-dependent speculation on
this issue can reduce the sensitivity to antigravity by less than an
order of magnitude.

The sensitivity suppression factor for a ${\overline{q}_1}{q_2}$
meson with mass $M$ can be estimated from the masses of current
quarks \cite{pdg} as
\begin{eqnarray}
\kappa_{\overline{q}_1{q}_2}&=&\frac{\Bigl\vert
m_{\overline{q}_1}-m_{q_2}\Bigr\vert}{m_M} \simeq
 \left\{
 \begin{array}{rcl}
 0.2\;,              &\mbox{for}& K\\
0.6\;,   &\mbox{for}& D\\
0.85\;,              &\mbox{for}& B
 \end{array}
 \right.\;.
 \end{eqnarray}

Another opportunity is a comparison of day's and night's results.
They may be different because of a different distance to the Sun.
The difference of potentials ranges from $\Delta U/c^2 \simeq
4.2\times10^{-13}$ to zero depending on the latitude. This level of
accuracy should clearly rule out antigravity from data on kaons and
$D$-mesons. The accuracy for $B$-mesons is at about a limit to
recognize whether the antigravity is possible.

Indeed, it may happen that experiments on the meson oscillations do
not allow a simple interpretation because it may be hard to
understand from the published results when exactly the data were
taken. However, in such a case the [anti]gravitational effects
should produce a `noise' at the level by a number of orders of
magnitude above the sensitivity and it should be seen.

One more option, which should work successfully for kaons and maybe
for $D$ mesons, is a comparison of data taken from different
laboratories. If their heights are different by 100 meters, the
difference in potentials is
\begin{equation}
\frac{\Delta U (\Delta r=100\;{\rm m}) }{c^2}\simeq
3.2\times10^{-14}\;.
\end{equation}

\section*{Conclusions}

We have demonstrated that a study of oscillations of neutral kaons
provides us with a precision test of the equivalence principle which
is different from other tests assuming CPT invariance and dealing
with specific gravitational sources and probe particles.

\begin{figure}[ptbh]
\begin{center}
\includegraphics[height=7cm,clip]{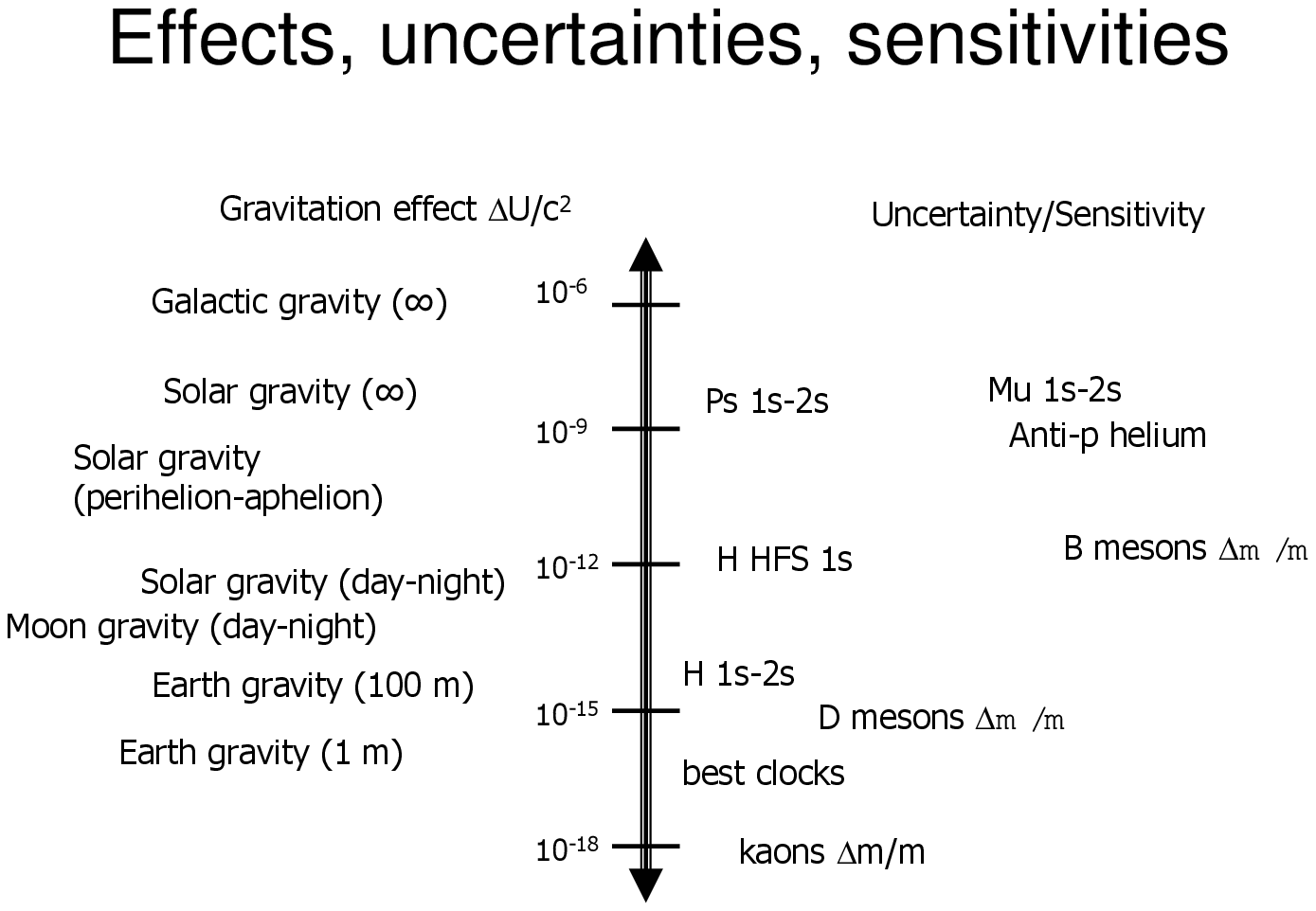}
\end{center}
\caption{Fractional values of gravitational effects versus
uncertainty and sensitivity of various precision measurements (see
also \cite{anti}).\label{f:sens} }
\end{figure}

There are a number of experiments with low uncertainty or with a
high sensitivity and a summary is presented in Fig.~\ref{f:sens}. It
happens that kaon physics cannot only provide the best constraint on
CPT invariance, but also offers a very sensitive tool to constrain
many other exotic options, such as, e.g., a departure from the
equivalence principle.

We have also proved that with no additional assumptions antigravity
should likely be in conflict with data on oscillations of neutral
mesons even in the gravitation field of the Sun and the Earth. More
accurate interpretation of the experiments may be needed for a
rigorous statement.

\section*{Acknowledgments}

The work was in part supported by DFG (under grant \# GZ 436 RUS
113/769/0-3) and RFBR (under grant \# 08-02-91969).

Stimulating discussions with S. I. Eidelman, R. A. Sunyaev, D. A.
Varshalovich, Th. Udem and V. G. Ivanov are gratefully acknowledged.

\end{document}